\documentclass[twocolumn,showpacs,preprintnumbers,amsmath,amssymb]{revtex4}

\usepackage{graphicx} 
\usepackage{dcolumn} 
\usepackage{bm} 

\begin{document}

\preprint{}

\title{Even-odd effect in Andreev Transport through a Carbon Nanotube Quantum Dot}
\author{A. Eichler}
\author{M. Weiss}
\author{S. Oberholzer}
\author{C. Sch\"onenberger}
\email{Christian.Schoenenberger@unibas.ch} \affiliation{Institut f\"ur Physik, Universit\"at Basel,
Klingelbergstr. 82, CH-4056 Basel, Switzerland}
\author{A. Levy Yeyati}
\author{J.  C. Cuevas}
\affiliation{Departamento de Fisica Teorica de la Materia Condensada,
Universidad Autonoma de Madrid, E-28049 Madrid, Spain.}
\date{\today}

\begin{abstract}

We have measured the current($I$)-voltage($V$) characteristics of
a single-wall carbon nanotube quantum dot coupled to
superconducting source and drain contacts in the intermediate
coupling regime. Whereas the enhanced differential conductance
$dI/dV$ due to the Kondo resonance is observed in the normal
state, this feature around zero bias voltage is absent in the
super\-conducting state. Nonetheless, a pronounced even-odd effect
appears at finite bias in the $dI/dV$ sub-gap structure caused by
Andreev reflection. The first-order Andreev peak appearing around
$V=\Delta/e$ is markedly enhanced in gate-voltage regions, in
which the charge state of the quantum dot is odd. This enhancement
is explained by a `hidden' Kondo resonance, pinned to one contact
only. A comparison with a single-impurity Anderson model, which is
solved numerically in a slave-boson mean\-field ansatz, yields good
agreement with the experiment.
\end{abstract}


\maketitle


There is a growing interest in the exploration of correlated
charge transport through nano\-scaled low-dimensional systems
involving both super\-conductors and normal
metals~\cite{Choi-Review,Ralph,Morpurgo,Glazman-Sdot,MAR-Atom-Contact,Review-Bouchiat}.
The penetration of the pair amplitude $\Delta$ from a super\-conductor (S) into a normal metal (N),
the proximity effect, is a manifestation of correlated charge transport
mediated by Andreev processes taking place at the S-N interface~\cite{Proximity-Effect} and
leading in S-N-S junctions to the Josephson effect~\cite{Josephson-Likharev}
and sup-gap current peaks due to multiple Andreev reflection (MAR)~\cite{Octavio}.
The super\-conducting proximity effect has been studied in great detail in
the mesoscopic size regime of diffusive, but phase coherent conductors~\cite{Meso-Supra}.
Andreev transport has also been the key quantity in experiments elucidating charge transport
in single atom contacts~\cite{MAR-Atom-Contact,MAR-Cuevas}.
On the other hand, Andreev transport through a quantum dot coupled
to super\-conductors, is just
emerging now~\cite{Buitelaar-SKondo,Buitelaar-MAR,Graeber-SKondo,Herrero-Jorgensen-Wernsdorfer}.
If the dot is weakly coupled to the leads, Andreev processes are suppressed by
the charging energy $U$ of the dot~\cite{Ralph,S-Resonance-S-Alfredo,S-Resonance-S-Wendin}.
If the dot is sufficiently small, size quantization takes place, forming a
quantum dot (QD) with discrete eigenstates (`levels') at energies $E_{\{i\}}$.
Transport then occurs through individual levels~\cite{Ralph}.
Since the level `positions' $E_{\{i\}}$, and sometimes also the coupling strengths
of the levels to both source and drain contacts $\Gamma_{1,2}$, can be tuned
through gate voltages, a physically tunable model system of
the Anderson `impurity problem' is realized.
With one electron on the QD (half-filling), a many-electron
ground-state forms, involving both the dot-state and
conduction electrons from the leads in an energy window given by the
the Kondo temperature $T_K$~\cite{Goldhaber-Gordon,Kondo-Effect}.
In this Kondo regime, which can be observed if $\Gamma_{1,2}$ is not too small,
a resonance pinned at the Fermi energy of the leads forms (Kondo resonance).
If superconducting contacts are used instead of normal ones,
the additional pair-correlation in the leads competes with the Kondo correlations
on the QD~\cite{Choi-Review,Glazman-Sdot,Schwab-SB-SKondo,Bergeret,Cuevas-Kondo-SN,Kondo-S}.
It has been found recently in experiments using carbon nanotubes (CNTs) as
QDs, that there is an interesting cross-over occurring at $k_BT_K \approx \Delta$.
If $\Delta > k_B T_K$, the Kondo correlations are suppressed, whereas they persist
in the opposite regime, opening a highly conducting channel for the
Josephson effect~\cite{Buitelaar-SKondo,Glazman-Sdot}.
CNTs are ideally suited for the realization of such systems, because CNTs can a)
act as well controlled QDs in different transport regimes~\cite{Review-Sapmaz},
including the Kondo regime~\cite{Kondo_CNT}, and b),
different types of contacts can be realized, including super\-conducting
ones~\cite{Morpurgo,Review-Bouchiat,Buitelaar-SKondo,Buitelaar-MAR,Graeber-SKondo,Herrero-Jorgensen-Wernsdorfer}.
Similar physics can be addressed with QDs fabricated in semiconducting
nanowires contacted to S contacts~\cite{Nanowires}.

We here report on measurements of the non-equilibrium (finite-bias) transport
through a single-wall carbon nanotube QD with S contacts
in the most interesting regime of intermediate coupling, where
Kondo correlations are of similar magnitude as superconducting ones.
We have found a pronounced even-odd effect in the MAR structure.


\begin{figure}
\includegraphics[width=85mm]{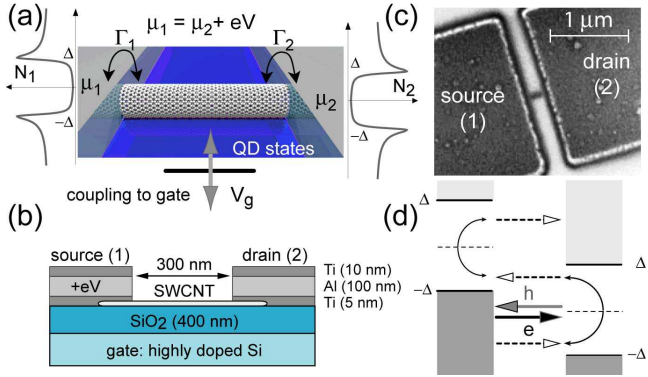}
\caption{\label{Fig1}
(a) Illustration of a SWCNT contacted by superconducting (Al)
source (1) and drain (2) electrodes. $\mu_{1,2}$ are respectively the electro\-chemical
potentials, $N_{1,2}$ the density-of-states, $\Gamma_{1,2}$ the life-time broadenings due to
the coupling of the QD state to the electrodes, $\Delta$ the super\-conducting gap parameter,
$V$ the applied source-drain voltage and $V_g$ the gate voltage.
(b) Device geometry, showing the evaporated tri-layer, consisting of a Ti adhesion layer, the super\-conducting
Al film and a Ti cap-layer.
(c) shows an actual device and (d) illustrates possible processes that lead to a subgap current.
Shown is in solid a first order Andreev process and dashed a second order one. In the first (second),
two (three) quasi\-particle (electrons $e$ and holes $h$) are involved.
} \end{figure}

Single walled carbon nanotubes were grown by chemical
vapor deposition on highly-doped
Si wafers with a \mbox{$400$\,nm} layer of thermal oxide on top,
using Fe particles as catalysts~\cite{1998-Kong,Comment-CVD}.
Individual single-wall carbon nanotubes (SWCNTs) were localized with a
scanning electron microscope and contacted to
super\-condcuting source and a drain electrodes
using e-beam lithography,
see Fig.~1.
%
The evaporated contacts consist of a Ti($5$nm)/Al($100$nm)/Ti($10$nm) tri-layer,
where Ti serves as an adhesion and cap-layer.
Al is the actual super\-conductor with a bulk critical temperature of
\mbox{$T_c=1.2$\,K}. In its thin film form, we rather measure
a $T_c$ of \mbox{$0.9$\,K}, which corresponds to a BCS gap-parameter
$\Delta_0=1.76 k_BT_c$ of \mbox{$0.135$\,meV}. We drive the Al contacts
into the normal state by applying a small perpendicular magnetic field
of \mbox{$B = 0.1$\,T}.
The substrate is contacted to a third terminal in order to
establish a backgate. We measure the differential source(1)-drain(2)
conductance $G:=dI/dV$ as a function of source-drain $V$ and gate-voltage $V_g$.
This is achieved by superposing an ac-voltage \mbox{$V_{ac}=10$\,$\mu$V}
on $V$ and measuring the corresponding ac current with a current-voltage
converter and a lock-in amplifier.
Several devices were fabricated and tested at room temperature and at \mbox{$4.2$\,K}.
Here, we focus on a particular interesting device which we selected and
measured in a dilution refrigerator.
This device has been studied over a large $V_g$
window and displays single-electron charging with addition
energies in the range of \mbox{$2\dots 5$\,meV}.
In the following we will focus on a confined gate-voltage regime.


\begin{figure}
\includegraphics[width=85mm]{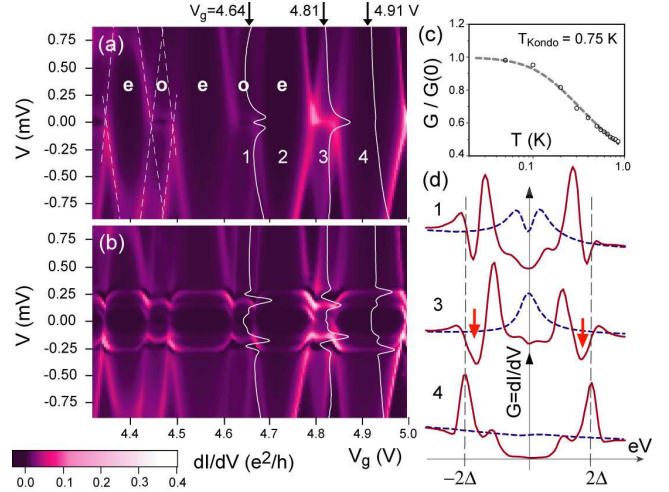}
\caption{\label{Fig2}
Differential conductance $dI/dV(V,V_g)$ plot as a function of bias $V$ and gate voltage $V_g$
of a SWCNT-QD with superconducting contacts in the normal (a) and super\-conducting state (b).
In (c) we show the linear conductance $G(T)$ measured as a function of temperature $T$
in the middle of charge state $3$. The curves in (d) correspond to the one overlaid on the
$dI/dV(V,V_g)$ plots.
} \end{figure}

Fig.~2 shows the main measurements we will be focusing on in the following.
Visible is in (a) a $dI/dV$ plot in the normal state (n-state) and the
corresponding one in the super\-conducting state (s-state) in (b).
In the n-state a sequence of larger and smaller Coulomb blockade (CB) diamonds
are seen (dashed lines), corresponding to a sequence of nearly equidistantly spaced
levels on the SWNT-QD, which are filled sequentially.
The number of electrons on the dot therefore alternates in the
ground-state between odd and even~\cite{CNTs-Odd-Even}.
It is also seen that the conductance $G=dI/dV$ around zero bias is suppressed
and featureless in the even valleys, but is increased assuming structure in the odd ones.
In the CB diamond labelled 3, there is a pronounced peak at $V=0$, suggesting
the appearance of a Kondo resonance.
Indeed, the dependence of the linear $G(T)$ on temperature $T$ (Fig.~2c),
measured in the middle of this valley,
follows the expected dependence~\cite{TK-of-T-dependence}
with a Kondo temperature of \mbox{$T_K=0.75$\,K}.
In the other odd valleys, the Kondo resonances are split by \mbox{$\approx 0.1$\,meV}~\cite{Comment1}.
The origin of this splitting is at present not known, but could be due to exchange with
ferromagnetic catalyst particles or another tube (if the present one is a
small bundle or contains more than one shell)~\cite{Kondo-Splitting}.

The n-state data can be used to deduce a number of parameters. The source, drain and gate capacitances
are \mbox{$C_{1,2}\sim 50,\,100$\,aF} and \mbox{$C_g\sim 4$\,aF}, leading to a gate-coupling
$\alpha=C_g/C_{\Sigma}$ of $\sim 0.026$, where $C_{\Sigma}=C_1+C_2+C_g$.
The charging energy $U=e^2/C_{\Sigma}$ and  the level spacing $\delta E$
are in the range of \mbox{$0.7\dots 1$\,meV} and \mbox{$1.4\dots 1.8$\,meV}, respectively.
Whereas this SWNT-QD is nearly symmetric in its electro\-static coupling,
it is quite asymmetric in its electronic one.
The total level broadening amounts to \mbox{$\Gamma=\Gamma_1+\Gamma_2 \approx 0.2$\,meV} with an
asymmetry of \mbox{$\Gamma_1/\Gamma_2\approx 50$}. This asymmetry is deduced from
the measured current peaks in $dI/dV$ at the border of the CB diamonds at finite bias and
is in agreement with the reduced
low temperature zero-bias $G(0)$ of the Kondo ridge 3, amounting to \mbox{$G(0)\sim 0.1$\,$e^2/h$}.

Looking next at the s-state, we see that the major changes in the $dI/dV$ are
confined to a voltage band of \mbox{$-0.26$\,meV\,$< V < 0.26$\,meV}, corresponding
to $\pm 2\Delta$. Above $2\Delta$, i.e. \mbox{$|V|>2\Delta/e$}, quasi\-particle
current is possible and the main modification is caused by the peak in the
super\-conducting density-of-state (DOS)~\cite{Ralph,MAR-Cuevas}, leading to a peak-like feature
in $dI/dV$. Due to the gap in the DOS, first order processes are however
suppressed below $2\Delta$. Depending on the transmission
probability~\cite{S-Resonance-S-Alfredo,S-Resonance-S-Wendin,Buitelaar-MAR},
higher order Andreev processes can contribute.
The first order Andreev process, for example, results in a peak-like structure
in the vicinity of $\Delta$. Due to the higher order, the first Andreev peak
and all subsequent ones are usually smaller than the quasi\-particle one.
Both the dominant $2\Delta$ and the smaller $\Delta $ peak, as well as the
suppressed $G$ in the s-state are best visible in the middle of
an even charge state (even valley), see e.g. curve labelled 4 in Fig.~2d.
In contrast, in the odd charge states, the $2\Delta$ feature
is not present or does not appear at $2\Delta$. Starting to view the data
from large bias voltage, the first peak appears closer to $\Delta$ rather than $2\Delta$, with
a preceding negative $dI/dV$ (NDR), see curve labelled 3 in Fig.~2d.
Hence, there is a striking even-odd asymmetry in the finite-bias $dI/dV$ features in the s-state
which is not caused by the CB resonance at the edges of adjacent charge states, where
the situation is expected to be more complex~\cite{S-Resonance-S-Alfredo,S-Resonance-S-Wendin}.
The even-odd alteration of the MAR structure suggest a relation to Kondo physics.
To model this, we first extract important parameters from an analysis of the data
in the middle of an even valley where Kondo correlations are absent.


\begin{figure}
\includegraphics[width=85mm]{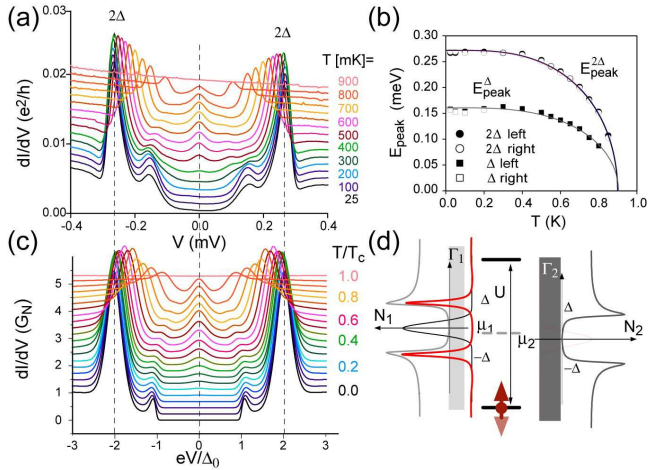}
\caption{\label{Fig3}
(a) measured $dI/dV$ versus temperature $T$ in the even valley 4 of Fig.~2a.
The $2\Delta$ and $\Delta$ peak positions $E_{peak}(T)$ are shown in
(b) together with a BCS $T$-dependence of $\Delta$.
In (c) we show the result of a model calculation based on tunneling between
two super\-conductors. Good agreement is obtained if the BCS-DOS is broadened
by an amount corresponding to $0.2\Delta_0$.
The illustration in (d) is our proposal to understand the appearance of the
strong $\Delta$ feature in the odd valleys. A Kondo resonance
persists on one electrode side only (here, drawn as a thin curve on the left side),
pinned to that chemical potential. Due to the gaped DOS, this resonance
splits leading to an enhancement of the $\Delta$ feature.
} \end{figure}

Fig.~3a-c discusses the temperature dependence of $dI/dV$ in the middle
of the even charge state; (a) shows the measurement taken in valley 4 of Fig.~2
and (c) is a model calculation.  The experiment displays pronounced
quasi\-particle current peaks at $E_{peak}=\pm 2\Delta$,
and weaker MAR peaks at $\pm \Delta$.
The evolution of $E_{peak}^{j\Delta}(T)$ with temperature $T$ are
shown in Fig.~3b together with an approximate BCS
gap function $E_{peak}^{j\Delta} = K_{j\Delta}\Delta_0\tanh(1.74 \sqrt{T_{c}/T-1})$,
where we used the BCS value for $\Delta_{0}=1.76k_{B}T_{c}$, which
amounts to \mbox{$0.135$\,meV} for a $T_c$ of \mbox{$0.9$\,K}. We then obtain
$K_{2\Delta}=2.0$ and $K_{\Delta}=1.15$ for the two peaks. The slight increase
of $K_{\Delta}$ above the expected value of $1$ is not unusual.
We will be using the value \mbox{$\Delta_0=0.13$\,meV} as the zero-T
gap parameter in the following. The relevant parameters expressed
in units of $\Delta_0$ are then:
$U=5\dots 8$, $\delta E = 10 \dots 14$, $\Gamma \approx 1.5$ and $T_K \approx 0.5$.
The zero-bias peak in Fig.~3a, appearing at intermediate temperatures,
can be explained by direct tunneling of quasi\-particles
thermally activated across the gap.

The good agreement with the BCS relation of the peak-positions
motivates the modelling of the $dI/dV$ using the BCS-DOS in the leads.
Although we would have to use a theory describing resonant tunneling between
two super\-conductors, such as the one from Levy Yayati {\it et al.}~\cite{S-Resonance-S-Alfredo},
a simple tunneling picture suffices~\cite{Tinkham}, because the
resonant levels are far away from the electro\-chemical potential of the
electrodes in the middle of a charge state.
The subgap current is treated in the same approximation using~\cite{MAR-Cuevas}.
To obtain a reasonable fit, the BCS-DOS has been convoluted with a Gaussian of width
$\eta$. A quite good agreeement is found with \mbox{$\eta=0.2$\,$\Delta_0$}.



We now turn our attention to the odd charge states.
We point out, that the zero-bias high-$G$ Kondo `ridge', which is associated with the Kondo resonance and
visible in the n-state, is not seen in the s-state. This is consistent with a Kondo temperature
$T_K$ that is smaller than $\Delta$, i.e. \mbox{$T_K=0.5\cdot\Delta$}~\cite{Glazman-Sdot,Buitelaar-SKondo}.
Although the Kondo resonance is not visible in the s-state, we suspect it to be responsible for
the even-odd asymmetry of the $\Delta$-feature in the s-state.

In the Kondo regime, the single spin on the QD in the odd
state is screened by exchange with conductance electrons from the leads. If the quasi\-particles
are bound in Cooper pairs in the leads, the Kondo temperature is renormalized, assuming a smaller value
$T_K^{\star}$. This renormalization is sensitive on the actual parameters $\Gamma_{1,2}$
and $\Delta$. Due to the asymmetry, it may happen that
a Kondo resonance with a reduced width forms on the contact with the larger $\Gamma$,
whereas on the other one it is suppressed. This is illustrated in Fig.~3d.

\begin{figure}
\includegraphics[width=85mm]{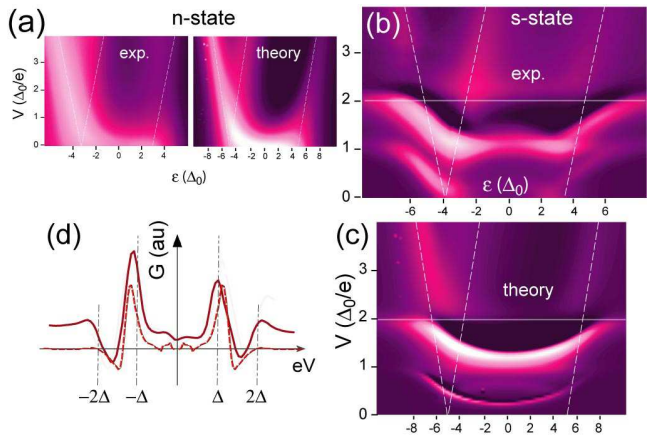}
\caption{\label{Fig4}
Comparison of the $dI/dV$ in the odd valley labelled 3 in Fig.~2a with a model caluclation
based a single-level Anderson model with interaction $U$, that is evaluated by a mean-field
slave-boson ansatz. The n-state is shown in (a), the s-state in (b) and (c).
$\epsilon$ denotes the level position.
The experiment corresponds to $U=7-8$ (in units of $\Delta_0$), whereas the calulation was
done for $U=5$ and $U=10$, where the latter is shown here.
The upper solid (dashed) curve in (d) is cross-sections in the s-state at \mbox{$\epsilon=0$},
taken from the experiment (theory).
} \end{figure}

We modelled this scenario considering a single-level Anderson Hamiltonian
with interaction $U=5\dots 10\,\Delta_0$ coupled to source and drain contacts.
The experimentally deduced $\Gamma$'s, including the strong asymmetry were used.
The calculation is based on a slave-boson mean-field treatment of this
interacting problem~\cite{Bergeret}. The result of the comparison is
shown in Fig.~4: (a) corresponds to the n-state and (b) and (c) to the s-state.
Despite this simple model, the agreement is surprisingly good. It is remarkably good in the
normal state, shown in Fig.~4a. In the s-state, the dominance of the $\Delta$-like feature
in the odd valley is clearly present, as is a similar cross-over from odd to
even filling. There are also some differences: in the experiment the $\Delta$-feature
bends to larger $V$-values in the middle of the odd state, whereas this feature
is rather flat in the calculation.


In conclusion we have discovered a pronounce even-odd effect in the
(multiple) Andreev structure in transport through
a QD with super\-conducting contact. This effect is  driven by
a Kondo resonance pinned to one contact only and defines a new
regime. Whereas a high conductance channel from source to drain,
driven by Kondo correlations, persists in the supercondcuting state
if $T_K \gg \Delta$, this channel is greatly suppressed in the opposite limit.
In the intermediate regime $T_K \sim \Delta$, and in particular
for asymmetric dot-electrode couplings, the (partial) Kondo-screening of the
`impurity' spin may occur on one electrode only.
It would be interesting to explore the `robustness' of this feature in model
calculation and to fabricate similar QDs with tunable electrode couplings.

\section{Acknowledgements}
We have profited from fruitful discussions with E. Scheer, W.
Belzig, and V. Golovach. We thank J. Gobrecht (PSI) for providing the
oxidized Si substrates. The work at Basel has been supported by
by the Swiss National Science Foundation, the NCCR on Nanoscale
Science, FP6-RTN DIENOW and EU-FP6-IST project HYSWITCH.


\begin{thebibliography}{99}
\bibitem{Comment1}
This splitting of the Kondo resonace is much larger than the
Zeeman splitting due to the applied magnetic field of
\mbox{$B=0.1$\,T}, amounting to \mbox{$< 10$\,$\mu$eV}.
\end{thebibliography}
\end{document}